# Spectro-Polarimetric Holographic Multiplexing Metasurface with Super-High Capacity Empowered by Mechanical Rotation

Ting Ma[1,2], Xianjin Liu[1,2], Qiwen Bao[1,2], Bolun Zhang[1,2] and Jun-Jun Xiao[1,2,*]

[1]College of Integrated Circuits, Harbin Institute of Technology (Shenzhen), Shenzhen 518055, China

[2]Shenzhen Engineering Laboratory of Aerospace Detection and Imaging, Harbin Institute of Technology (Shenzhen), Shenzhen 518055, China

E-mail: eiexiao@hit.edu.cn

**Abstract:** Mechanically reconfigurable metasurfaces capable of translation, rotation, and permutation have attracted considerable attention for high-capacity optical information storage and full-color holographic displays, owing to their low-power and high functional scalability, despite the additional system-level complexity introduced by precision rotation stages. This study presents a differentiable inverse design framework for such metasurfaces, creating an accurate mapping between meta-atom geometries and their multi-channel optical responses across diverse optical dimensions. Using a deep neural network-driven, end-to-end optimization pipeline, the framework enables intelligent, iterative refinement of rotatable metasurface within constrained design space. Using this approach, we show high-fidelity holographic video display by rotating a single element in a cascaded metasurface doublet around the optical axis. The doublet enables pixel-resolved holographic imaging with 288 independent channels, and by switching input/output polarization states, the system demonstrates four distinct full-color dynamic holographic videos. This work establishes an alternative paradigm for optical parameter multiplexing and end-to-end inverse design in mechanically reconfigurable metasurfaces, suggesting applications in compact optical systems, dynamic holography, information processing, and optical computing.

# 1. Introduction

Holographic optical data storage enables the direct reconstruction of the physical light waves of objects in a target scene, providing a natural naked-eye 3D viewing method. Consequently, it is considered one of the ultimate 3D display technology solutions, overcoming the vergence-accommodation conflict inherent in traditional stereoscopic display technologies.[1] The core advantage of this technology lies in its ability to fully record and reproduce multidimensional optical information such as amplitude, phase, and polarization of object light waves. Metasurfaces can achieve precise control over multiple optical parameters of light fronts, including phase, amplitude, polarization state, and orbital angular momentum, through parameter design (material, geometric configuration, arrangement, etc.) of artificial microstructures at sub-wavelength scales, thereby surpassing the limitations of traditional optics.[2] Metasurfaces have facilitated the miniaturization and integration of optical devices, leading to breakthroughs in fields such as optical imaging,[3–5] sensing,[6,7] communication,[8,9] and information processing,[10,11] optical computing[12,13] and holographic optical storage.[14,15] With the exponential growth in demand for data throughput in the information society, enhancing the channel capacity of optical systems has become a key issue.

As shown in **Figure 1(a)**, a majority of current studies primarily focus on three aspects of the multiplexing strategies of metasurfaces and system capacity expansion: (1) multiplexing in the dimension of optical parameters, (2) electrical domain reconfiguration multiplexing, and (3) mechanical reconfiguration multiplexing. It is well known that light fields are typically characterized by a bunch of optical parameters including amplitude, phase, polarization, wavelength, K-vector propagation direction, orbital angular momentum, and independent multiplexing channels over time.[16] Optical multiplexing over these parameter dimensions refers to transmitting different light field information by independently manipulating the various relevant optical parameters, such as wavelength,[17–19] polarization (for example, with total channel numbers of 1,[20] 2,[21] 3,[22] 4,[23] 6,[24] 11,[25] 55,[26]) propagation direction,[27] diffraction distance,[28] incident angle,[29] and orbital angular momentum.[30] This allows different optical properties to coexist and operate independently, significantly enhancing the capacity and efficiency of optical communication and information processing while expanding channel capacity. However, the resultant improvement in channel capacity remains markedly restricted due to the heightened inter-channel correlation among metasurface channels, which exacerbates crosstalk interference during optical parameter multiplexing operations.

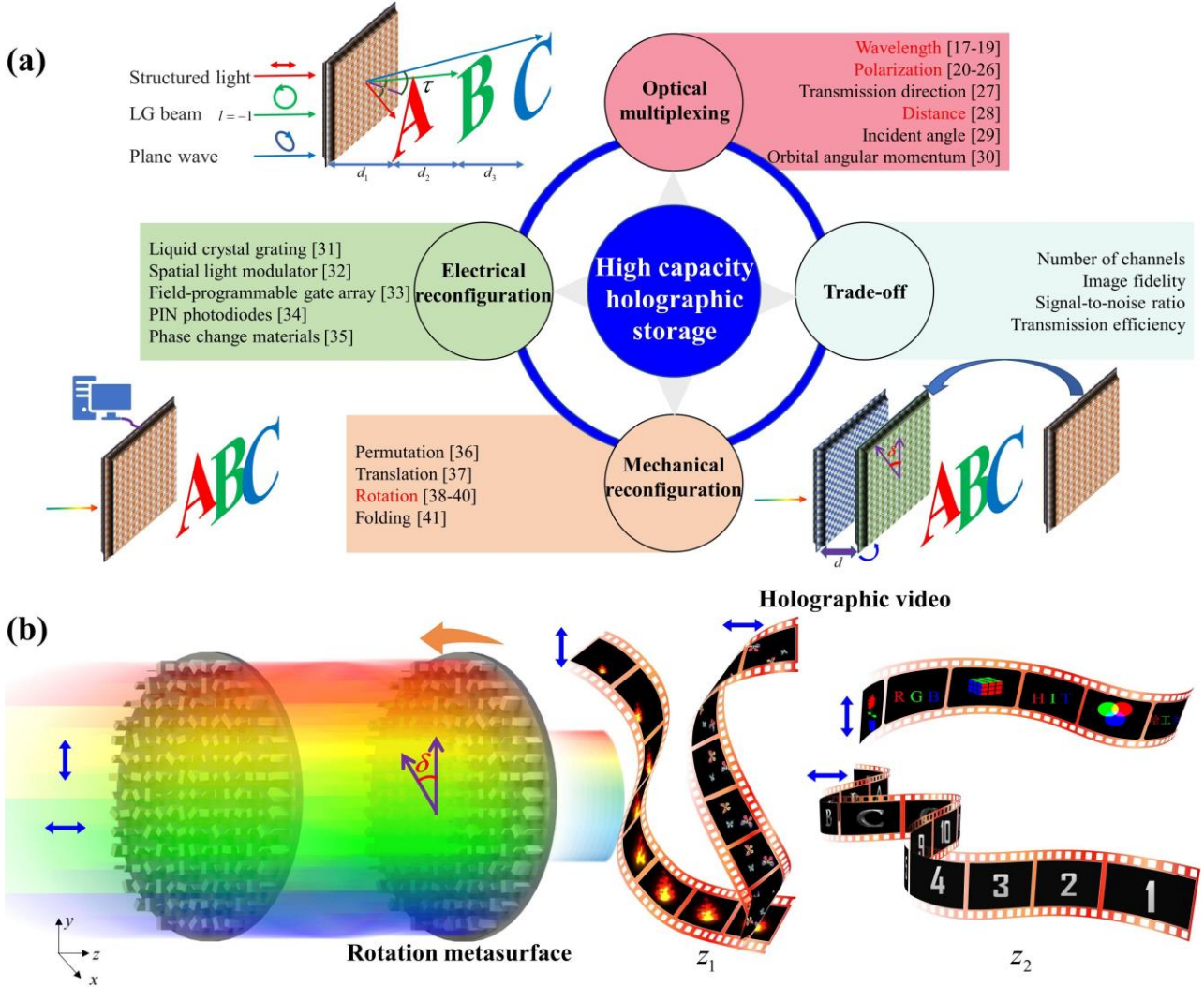

**Figure 1.** High-capacity holographic using multiplexed metasurface. (a) Three methods for expanding the storage capacity of metasurfaces as reported in the literature: are optical parameter multiplexing, electrical domain reconfiguration, and mechanical reconfiguration. This work employs a combination of mechanical reconfiguration and optical parameter multiplexing specifically highlighted in red. (b) Schematic diagram of the proposed high-capacity metasurface holographic video. Note that the video frame is governed by the rotation of one metasurface and the selection of the four videos depends on the incident and detection polarization (see blue arrows).

Dynamic metasurfaces, due to their reconfigurable or programmable nature, have also been widely applied to improve system capacity. To achieve dynamic control of the optical response of materials, common strategies for electrically reconfigurable metasurfaces involve introducing external stimuli through the design of active metasurfaces, such as liquid crystal gratings,[31] spatial light modulators,[32] and field-programmable gate arrays.[33] However, these methods usually require integrating additional electronic controls within the meta-atoms, such as PIN diodes[34] and phase-change materials like $VO_2$,[35] which not only increase the complexity and size of the metasurfaces but also lead to higher energy consumption.

Mechanically reconfigurable metasurfaces, however, provide a low-power solution for dynamically manipulating light by altering the metasurface's domain morphology concerning the incident light field, through mechanical operations. This approach offers relatively low complexity in fabrication compared to electrically or optically reconfigurable metasurfaces. Since each meta-atom undergoes limited shape changes during a single mechanical operation, reconfiguring a layer in cascaded metasurfaces through permutation,[36] translation,[37] rotation,[38–40] and folding[41] offers a more flexible approach. However, the number of channels reported remains limited due to the inefficiency of existing optimization models, which fall short of meeting the demands of practical applications. Additionally, there is a significant trade-off between the number of channels in holograms and fidelity, signal-to-noise ratio, and transmission efficiency. As the number of channels increases, systems often need to moderately reduce transmission efficiency to maintain necessary imaging clarity. Such parameter optimization strategy has proven to be an effective compromise.[42] Under the constraint of fixed model parameter scales, it is possible to break the channel capacity limitations, by making a trade-off between the performance (quality) and the overall capacity.

Compared with amplitude-only holography and phase-only holography, complex amplitude holography demonstrates significant advantages in terms of reconstruction fidelity.[43] However, in the transmission response model of meta-atoms established based on electromagnetic simulation methods, for example, finite-difference time-domain (FDTD), rigorous coupled-wave analysis (RCWA), and finite-element method, the geometric parameters and complex amplitude distribution are in a non-differentiable mapping relationship. This severely restricts the control capability of multi-channel complex amplitude holographic display across multiple optical parameters. In the multi-channel optimization design of metasurfaces based on the adjoint method, two full-wave numerical simulations are required in each iteration cycle to obtain the parameter perturbation gradient. This, however, significantly increases the computational resource consumption and computation time, causing the optimization efficiency to decrease exponentially.[44] It is worth noting that, based on the modeling capability of deep learning for complex non-linear mapping relationships, the geometric layout of metasurfaces can be inversely designed in an end-to-end manner to improve the design efficiency of multi-channel complex amplitude light field modulation of metasurfaces.[45–47] By optimizing phase profiles across multilayer diffractive optical elements, diffractive optical neural networks (DONNs) achieve multi-objective mappings.[48–50] We extend this framework to multichannel applications through a DONN-based computational architecture that enables computationally efficient inverse design of multilayer metasurface devices.

In this regard, we introduce a metasurface-based differentiable diffractive optical neural network (meta-diff-DONN) framework, which realizes end-to-end optimization of multi-layer metasurfaces by constructing a differentiable forward model. We use deep learning to build a surrogate model of the electromagnetic solver, effectively fitting the non-linear mapping relationship between the geometric parameters of the metasurface and the complex transmission coefficient. Driven by the gradient descent algorithm, we can iteratively update the metasurface structural parameters to achieve precise matching of the target light field automatically and stably. First, a deep neural network (DNN) is used to fit the complex mapping relationship between the metasurface geometric layout and its complex amplitude, serving as a surrogate model for the electromagnetic solver. The surrogate model of the electromagnetic solver is embedded in the inverse design chain to realize the multi-functional end-to-end design of metasurfaces and guide the multi-objective system to find the optimal solution. This study introduces a novel concept of mechano-optical hybrid reconstruction. By integrating the inherent optical multiplexing characteristics of metasurfaces with mechanically reconfigurable mechanisms, a composite design space involving rotational operations and three sets of optical parameters (wavelength, polarization, and diffraction distance) is constructed. Consequently, our meta-diff-DONN establishes a multi-dimensional space with thousands of potential interdependent channels through this integrated design framework. The details as schematically shown in Figure 1(b). This work achieves a channel capacity far beyond that enabled by conventional wavelength-polarization multiplexing techniques. We propose a dual-layer metasurface system that facilitates spectro-polarimetric-rotational holography. The decoupled design freedoms of each metasurface layer provide unprecedented control over multiplexed image assignment. In short, this study establishes a new design paradigm in metasurface holography—enabling 288 holographic channels within a single device footprint. This technology unlocks new application pathways in ultra-high-density optical storage, multi-user AR/VR displays, and dynamic optical encryption.

## 2 Meta-diff-DONN end-to-end inverse design framework

Let's first establish a physical model for the multi-channel multiplexing metasurface shown in Figure 1(b). Our meta-diff-DONN, as illustrated in **Figure 2(a)**, requires an end-to-end inverse design to identify the optimal geometric parameter layout of the metasurface for comprehensive parametric control over spatial light waves. This task is formulated as a multi-objective optimization problem

$$\arg\min_{\mathbf{G}} Loss = \arg\min_{\mathbf{G}} \sum_{\lambda,\sigma,z} F[U_{out}(\lambda,\sigma,z;\mathbf{G}), U_{tar}(\lambda,\sigma,z)], \qquad (1)$$

where $U_{tar}$ represents the target spatial light field characterized by amplitude $(A)$, phase $(\varphi)$, wavelength $(\lambda)$, and polarization $(\sigma)$ at each spatial position $(x,y,z)$ and $U_{out}$ denotes the field modulated by the metasurface with finite geometric parameters $\mathbf{G} \in [\mathbf{G}_{min}, \mathbf{G}_{max}]$. We use the mean squared error (MSE) and negative Pearson correlation coefficient (NPCC) of the light intensity $I = |U|^2$ as the loss function to adjust the transmission light field of the metasurface. Details are provided in Supporting Information Note S1.

This high-dimensional optimization problem constrained by multiple objectives is solved iteratively by minimizing the loss function using gradient descent-based methods. The forward model of this process includes complex amplitude modulation by two layers of metasurfaces and two diffraction propagation processes. The second layer of the metasurface can be rotated as a whole to obtain more diffraction patterns. The diffraction propagation process is calculated using the angular spectrum method. The geometric layout of meta-atoms determines the corresponding complex amplitude modulation under incident light fields $U_{in}(\lambda,\sigma)$, which is obtained via a neural network-assisted dispersive full-parameter Jones matrix, as shown below in Figure 2(a). Each meta-atom consists of a single titanium dioxide ($TiO_2$) rectangular unit on a silica ($SiO_2$) substrate, featuring sub-wavelength square lattice constant $P = 400$ nm, fixed height $H = 600$ nm, and three variable geometric parameters: length $(L)$, width $(W)$, and orientation angle $(\theta)$. $L$ and $W$ both vary from 50 to 250 nm and $\theta$ varies from 0 to $2\pi$. The metasurface is defined by the set of geometric parameters $\mathbf{G}$ of meta-atoms at coordinates $(x, y)$, denoted as $\mathbf{G} = \{L, W, \theta \mid (x, y)\}$.

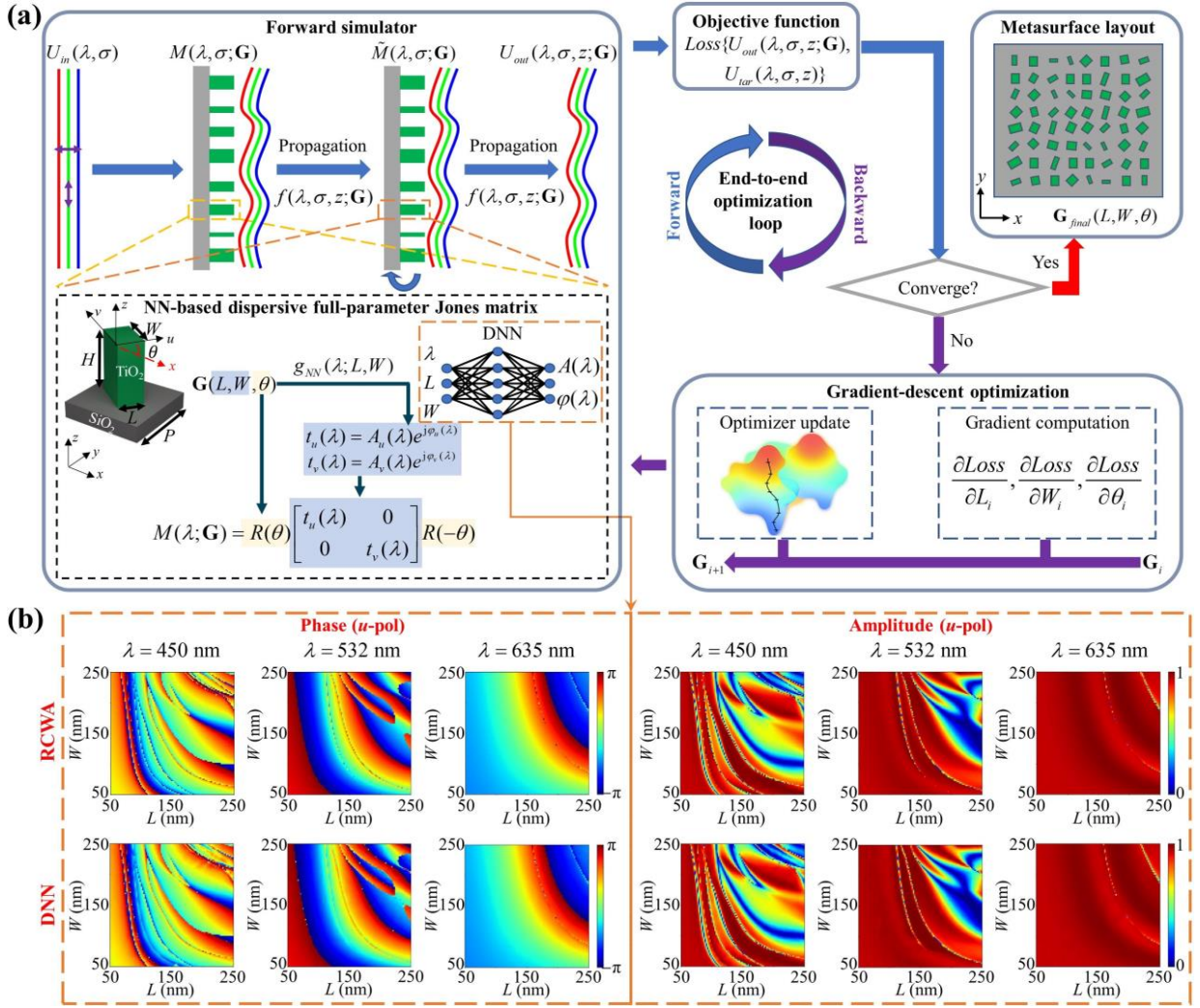

**Figure 2.** Meta-diff-DONN multi-channel multiplexing design routine. (a) Meta-diff-DONN end-to-end inverse design optimization loop. (b) Comparison of the RCWA simulation and DNN prediction for the amplitude $A_u$ and phase $\varphi_u$, at three wavelengths and $u$-pol, nanoblock meta-atoms with orientation angle $\theta = 0$.

However, the direct electromagnetic numerical simulations to establish the complex mapping between meta-atom geometries and their corresponding transmission coefficients prove both time-consuming and non-differentiable. To address this challenge, we employ a DNN as a surrogate model for the electromagnetic solvers, establishing the relationship between metasurface geometric parameters $(L,W)$ and their amplitude and phase responses under multiple wavelengths and polarizations. The first row of Figure 2(b) illustrates the dataset required for constructing this mapping relationship, obtained through numerical simulations using the RCWA method. For RGB trichromatic $u$-polarized $(\theta=0, u=x)$ light at the wavelengths of $\lambda_1 = 450$ nm, $\lambda_2 = 532$ nm, $\lambda_3 = 635$ nm, we have calculated the transmission coefficients of rectangular meta-atoms under varying length and width dimensions via the RCWA, subsequently deriving corresponding amplitude and phase responses. A three-layer fully connected neural network was trained to establish this mapping relationship. As shown in the second row of Figure 2(b), the DNN-predicted amplitude and phase demonstrate excellent agreement with the ground truth results obtained by the RCWA approach, achieving a complex transmission coefficient fitting error of 0.0063 calculated by MSE across three wavelengths. This minimal discrepancy confirms the successful training of the DNN. Detailed training protocols are provided in Supporting Information Note S2.

Upon completion of training, the stabilized DNN weights enable rapid generation of optical phase and amplitude responses from geometric configurations through its analytical expression $A_u, A_v, \varphi_u, \varphi_v = g_{NN}(\lambda; L, W)$. The dispersion and polarization characteristics of the meta-atoms at each position $(x, y)$ on the metasurface can be represented by a dispersion-enabled full-parameter Jones matrix:

$$M(\lambda; \mathbf{G}) = R(\theta) M_0(\lambda; L, W) R(-\theta) = \begin{bmatrix} t_{xx} & t_{xy} \\ t_{yx} & t_{yy} \end{bmatrix}, \qquad (2)$$

where the rotation matrix $R(\theta) = \begin{bmatrix} \cos\theta & -\sin\theta \\ \sin\theta & \cos\theta \end{bmatrix}$ accounts for geometric orientation, and the orientation angle $(\theta)$ of the meta-atom is positive in the counterclockwise direction. When the nanoblocks remain unrotated (orientation angle $\theta = 0$), the Jones matrix is represented as $M_0(\lambda; L, W) = \begin{bmatrix} t_u(\lambda; L, W) & 0 \\ 0 & t_v(\lambda; L, W) \end{bmatrix}$, while $t_u(\lambda; L, W)$ and $t_v(\lambda; L, W)$ respectively denote wavelength $(\lambda)$ and geometry $(L,W)$-dependent transmission coefficients along the fast (long) and slow (short) axes that have been analytically modeled by the well-trained DNN. With meta-atoms initially unrotated $(\theta = 0)$, the four complex components of this wavelength-dependent Jones matrix explicitly manifest dispersion control capabilities.

Note that complex diffraction within the metasurface generates unique optical responses. Continuous rotation of the metasurface induces the sequential emergence of new diffraction domains $D_1, D_2, ..., D_k$ at angular increments $\delta_1, \delta_2, \delta_3, ..., \delta_k$, enabling holographic video reconstruction through the dynamic selection of stored frames mapped to specific rotation angles. The secondary metasurface layer introduces rotational degrees of freedom, with its modulated capability described by

$$\tilde{M}(\lambda; \mathbf{G}(\tilde{x}, \tilde{y})) = R(\delta + \theta) M(\lambda; \mathbf{G}(x, y)) R(-(\delta + \theta)), \quad (3)$$

where $(\tilde{x}, \tilde{y})$ is the new position coordinate of the meta-atom after the overall rotation of the metasurface, and $(x, y)$ is the original coordinate of the meta-atom on the metasurface. Considering the inherent discreteness of the data, coordinate alignment during rotation operations was facilitated through bi-linear interpolation.[14] Despite this, slight residual errors may remain. Therefore, adopting a polar coordinate system to define the subwavelength unit cell grids rather than a Cartesian one could be of advantage. A comprehensive discussion is provided in Supporting Information Note S3. Considering incident fields $U_{in}(\lambda, \sigma)$, the modulated fields $U_{out}(\lambda, \sigma)$ after interaction with the differentiable neural network-based Jones matrix undergo wavelength- and polarization-specific diffraction at designated regions $(x, y, z)$. The differentiable forward model of the single metasurface can be shortly summarized as

$$U_{out}(\lambda, \sigma; (x, y, z)) = f_{propaga.} \left\langle M(g_{NN}[\lambda, L(x,y), W(x,y)], \theta(x,y)) \bullet U_{in}(\lambda, \sigma), z \right\rangle. \quad (4)$$

The initial geometric configuration $\mathbf{G}(L, W, \theta)$ is randomly generated, with the angular spectrum method $(f_{propaga.})$ employed for forward propagation calculations (Supporting Information Note S4). Rotational manipulation of the secondary metasurface layer generates twin diffraction domains corresponding to preset angles, producing mapped image sets at the output. This forward simulator enables automatic differentiation to extract gradient information $\partial Loss / \partial \mathbf{G}$ through the loss function.

Subsequent parametric optimization employs the Adam optimizer to update geometric parameters $\mathbf{G}$. The automatic differentiation-based gradient backpropagation is implemented using the PyTorch framework. Iterative execution of forward propagation and error backpropagation processes continues until design criteria are met, ultimately encoding the complete image sets into the metasurface and outputting the final geometric parameters $\mathbf{G}_{final}(L, W, \theta)$ as shown in Figure 2(a).

## 3 Results and discussions
### 3.1 Wavelength, polarization, and diffraction distance multiplexing

We first use our end-to-end inverse design method to achieve wavelength, polarization, and diffraction distance multiplexing holography via a single metasurface, without considering its overall rotation. As shown in **Figure 3(a)**, we examine a metasurface of size $200 \times 200$ μm$^2$, with a total of $500 \times 500$ units. We integrate two imaging planes with different diffraction distances, located respectively at $z_1 = 300$ μm and $z_2 = 500$ μm. We use three wavelength channels for color holography. As indicated in Equation (2), the dispersive full-parameter Jones matrix has three wavelength-dependent parts: $t_{xx}(\lambda), t_{yy}(\lambda)$, and $t_{xy}(\lambda) = t_{yx}(\lambda)$. [25] These three parts can be utilized by alternatively switching the input and output linear polarization. Details of the polarization conversion are in Supporting Information Note S5.

When the incident and output light are both *x*-polarized, color holograms of "Flower" and English letters "RGB" are generated on the depth planes $z_1$ and $z_2$, respectively. When the incident and output light are *x*-polarized and *y*-polarized, respectively, the meta-optical device shows color holograms of "Magic cube" and "HIT" on the depth planes $z_1$ and $z_2$, respectively. When the incident and output light are both *y*-polarized, color holograms of "RGB palette" and Chinese characters "哈工大" are generated on the depth planes $z_1$ and $z_2$, respectively. Figure 3(b) shows the length $(L)$, width $(W)$, and orientation angle $(\theta)$ distribution maps of the meta-atoms of the optimized metasurface. As shown in Figure 3(c), the reconstructed images on each *z*-plane are very similar to the target holograms. Figure 3(d) shows the observation results of each independent channel in Figure 3(c), with color, polarization, and distance channels separated for better clarity. To assess hologram quality, we use the structural similarity index measure (SSIM) to evaluate the similarity between the output hologram and the target image, and the peak signal-to-noise ratio (PSNR) to analyze the noise intensity in the hologram. Details of image quality evaluation indicators are in Supporting Information Note S6. The PSNR of each hologram is shown in the lower-left corner. The average PSNR and SSIM of the 18-channel holographic image are 26.09 dB and 0.72, respectively. Moreover, we verify the reliability of the metasurface design through full-wave simulations using the FDTD method. Note that considering the limits of computing memory and speed, the metasurface size and diffraction distance are scaled down to $100 \times 100$ pixels and 45 μm, respectively. For this size-reduced metasurface, the length $(L)$, width $(W)$, and orientation angle $(\theta)$ distribution maps are independently optimized (details in Supporting Information Note S7). The 3D full-wave simulation results verify the outputs of the physical model. To enhance the fabrication-error robustness of holographic metasurfaces, we introduced Gaussian noise representing shape variations in meta-atoms during the end-to-end optimization process, thereby realizing a defect-tolerant metasurface design. The holographic images maintained consistent quality under meta-atom shape variations of ±20 nm without significant degradation. Detailed results are presented in Supporting Information Note S8.

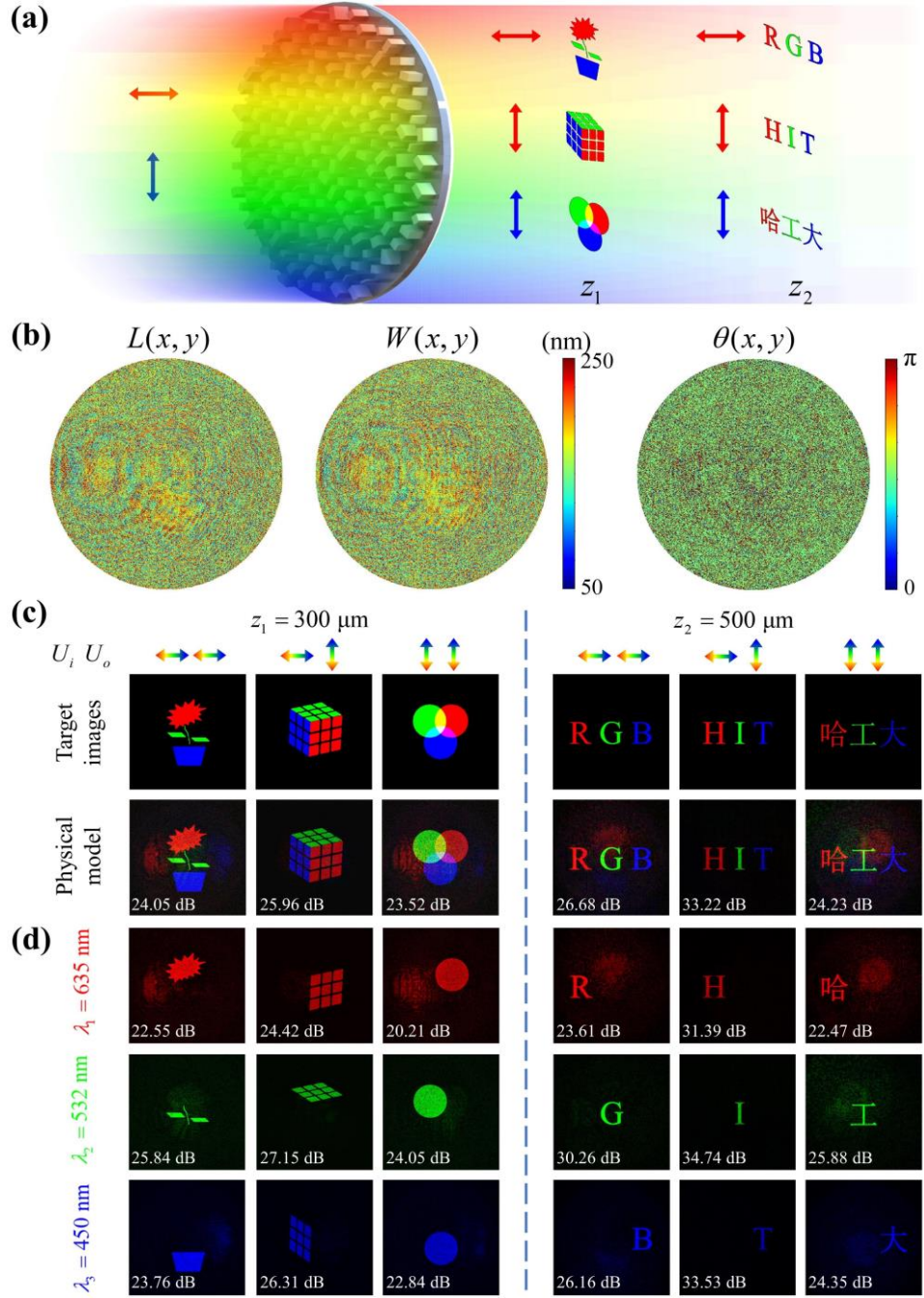

**Figure 3.** Three-polarization color 3D holographic display. (a) Schematic diagram of wavelength, polarization, and diffraction distance multiplexing. (b) Distribution map of length ($L$), width ($W$), and orientation ($\theta$) of meta-atoms for the inverse-designed metasurface. (c) Under different input/output polarizations, the target and physical model output color holographic images of "Flower" "Magic cube" and "RGB palette" at the depth plane $z_1$, while the target and physical simulation color holographic images of the letters "RGB", "HIT" and the Chinese characters "哈工大" at the depth plane $z_2$. (d) Single holographic images output from the 18 holographic channels of the metasurface, with the PSNR of each holographic image relative to the target image labeled at the bottom left corner of the figure.

**3.2 Polarization and rotation angle multiplexing**

In this section, we demonstrate the mechanical rotation of one metasurface inside a metasurface doublet. A bilayer metasurface configuration is set where the first metasurface remains fixed while the second metasurface rotates around the optical axis at multiple angles. The interaction between these two metasurfaces generates novel light field modulation. According to Rayleigh-Sommerfeld diffraction theory, each meta-atom on the metasurfaces functions as an independent secondary source, influencing subsequent layers through distinct diffraction domains labeled as $D_0$. When applying a rotational/twist angle $\delta$ to the second metasurface, the diffraction domain maintains consistent morphology but exhibits switched effective interaction regions. As the twist angle varies from $\delta_1, \delta_2, \delta_3, ..., \delta_k$, the system transitions from the initial diffraction domain $D_0$ to a new set of diffraction domains $D_1, D_2, D_3, ..., D_k$, which introduces additional design freedom despite the shared diffraction domains. Continuous rotation of the metasurface enables dynamic refreshment of optical responses and projection of holographic videos encoded in the metasurfaces. By mapping temporal information of original videos to the discrete twist angles, we achieve frame-selective holographic video display through twist-dependent dynamic selection.

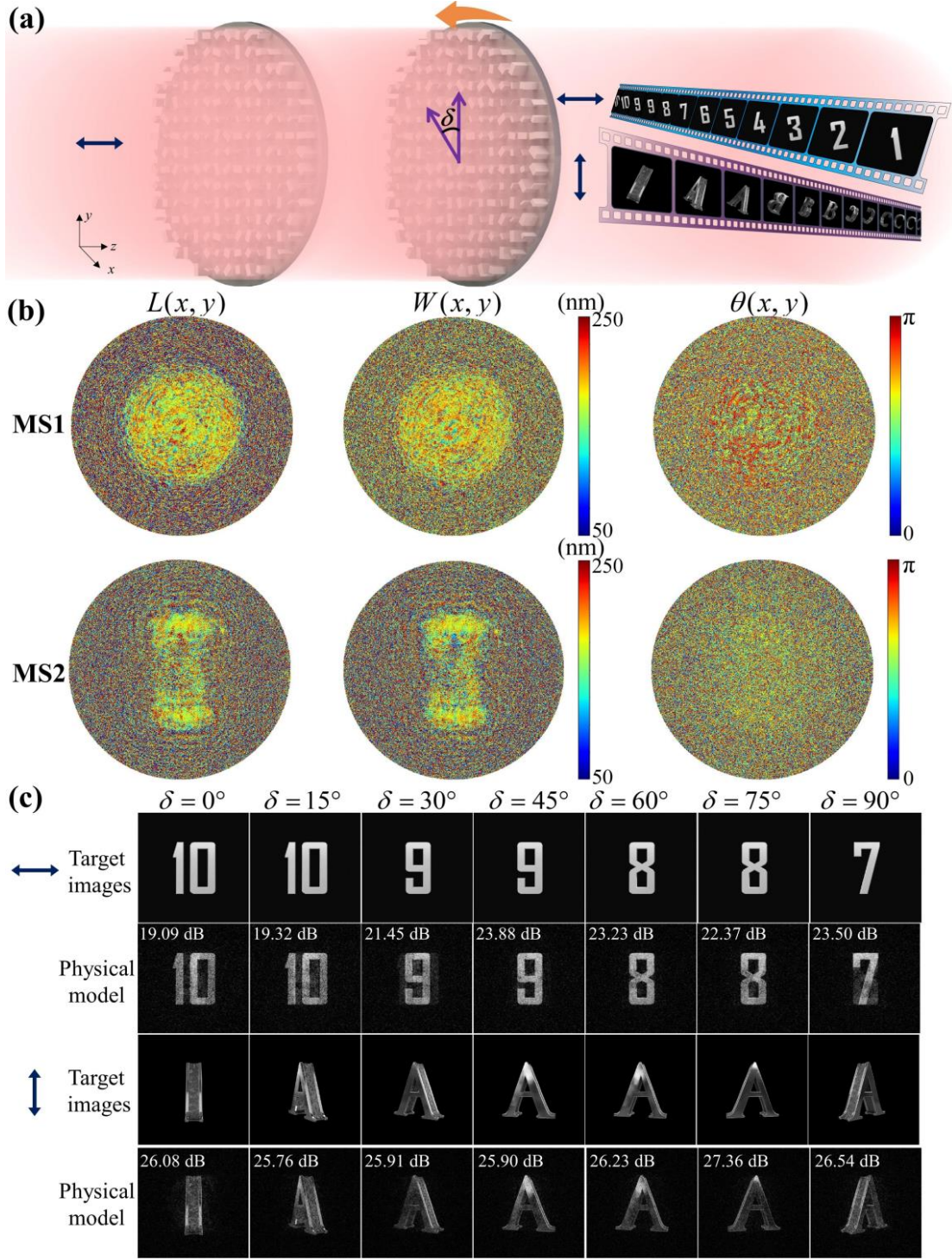

**Figure 4.** Dual-detection-polarization holographic display by rotating one metasurface in a doublet. (a) Schematic diagram of polarization and rotation holographic multiplexing. (b) Inverse design of the length ($L$), width ($W$), and angle of orientation ($\theta$) distribution of the metasurface doublet. MS is the abbreviation for metasurface. (c) Under different output polarization, the target and physical model output monochromatic holographic images of "Countdown" and "ABC", with the PSNR of each holographic image labeled in the upper left corner of the panels.

**Figure 4(a)** shows the scheme of polarization and rotation-angle multiplexed metasurface. Each layer comprises a circular metasurface with $500 \times 500$ meta-atoms and meta-atom pitch $P = 400$ nm, designed for wavelength $\lambda = 635$ nm. Note that the diffraction distance and the metasurface separations are both $z = 230$ μm, and 24 rotational angles $\delta$ at 15° intervals for a full 360° rotation are implemented. When simulating free-space optical propagation using the angular spectrum algorithm, zero-padding around the optical field is essential to prevent spectral aliasing.[51] However, this zero-padding substantially enlarges the computational grid size. For given computational resources, a trade-off must be made between diffraction distance and training batch size to improve the stability and convergence speed of the metasurface inverse optimization. Under *x*-polarized illumination, continuous rotation at constant speed projects 24 distinct monochromatic images (equivalent to 24 fps) of the "Countdown" animation onto the *x*-polarized output detector. Switching to *y*-polarized detection enables the synchronous projection of the "ABC" sequence. See Supporting Video 1 demonstrates dual-video output. Figure 4(b) presents optimized geometric parameters $(L, W, \theta)$ of all meta-atoms. Figure 4(c) displays a portion of the holographic output sequences at different rotation angles for both *x*- and *y*-polarization detection channels, with full 48-frame sequences. The training details are provided in Supporting Information Note S9. The quantitative evaluation shows an average PSNR of 23.36 dB (SSIM of 0.71) for *x*-polarized outputs and 26.48 dB PSNR (SSIM of 0.73) for *y*-polarized outputs. These results clearly show that we have successfully constructed high-quality $500 \times 500$ monochrome videos through controlled metasurface rotation at designated output planes.

**3.3 Wavelength, polarization, and rotation angle multiplexing co-design**

To show the model's capability in more multidimensional channel control scenarios, we employ the meta-diff-DONN framework to co-design wavelength-, polarization-, and rotation-angle multiplexed holography. Multiple video frames are encoded into a bilayer metasurface architecture, selectively activated by *x*- or *y*-linear polarized illumination. Temporal information of original videos is mapped to discrete rotation angles, enabling dynamic frame selection through metasurface rotation for holographic video reconstruction of stored color sequences. Our meta-diff-DONN model is implemented using PyTorch 1.10.1 on a workstation equipped with a Tesla V100 GPU (Nvidia Inc., 32GB VRAM) and two Intel Xeon Gold 5115 CPUs (256GB RAM). Due to the limitation of computing resources, we train the 288-channel images ($500 \times 500$ pixels) in batches and set batch size $B=12$. The wall-clock time to reach convergence is approximately 34.05 hours. **Figure 5(a)** schematically illustrates the wavelength-polarization-rotation multiplexed holographic architecture. By switching input/output polarization states, the metasurface projects four distinct color videos: 1) "Flame" under *x*-polarized RGB illumination with *x*-polarized detection; 2) "Yellow flower" with *x*-to-*y* polarization conversion; 3) "Colorful butterfly" under *y*-polarized RGB illumination with *x*-polarized output; 4) "Blue butterfly" under *y*-to-*y* polarization configuration (Supporting Video 2). The dual-layer metasurface composed of discrete nanoblocks achieves decoupling of the diagonal elements in the Jones matrix, thereby establishing four independent polarization conversion channels through its anisotropic architecture, see more details in Supporting Information Note S3. Figure 5(b) displays the first eight holographic frames of each sequence, with complete sequences provided in Supporting Information Note S10. Notably, the "Flame" sequence exhibits relatively lower image quality, likely due to its complex spatial features. Results demonstrate that merely two metasurface layers can effectively store and retrieve up to $3 \times 4 \times 24 = 288$ relatively high-quality images. It is important to note that the image quality could be improved by balancing the multiplexed channel number and the total meta-atom design space size. The validity of the bilayer metasurface simulations was further corroborated using small-scale FDTD simulations, as demonstrated in Figure S14. While this implementation case utilizes wavelength, polarization, and global rotation multiplexing, incorporating diffraction-distance multiplexing could further expand the channel capacity. However, increased channel numbers would inevitably introduce elevated background noise and reduced signal-to-noise ratio (SNR)—a trade-off arising from target holographic channels while maintaining identical meta-atom design space $\mathbf{G}(L, W, \theta)$ for encoding.

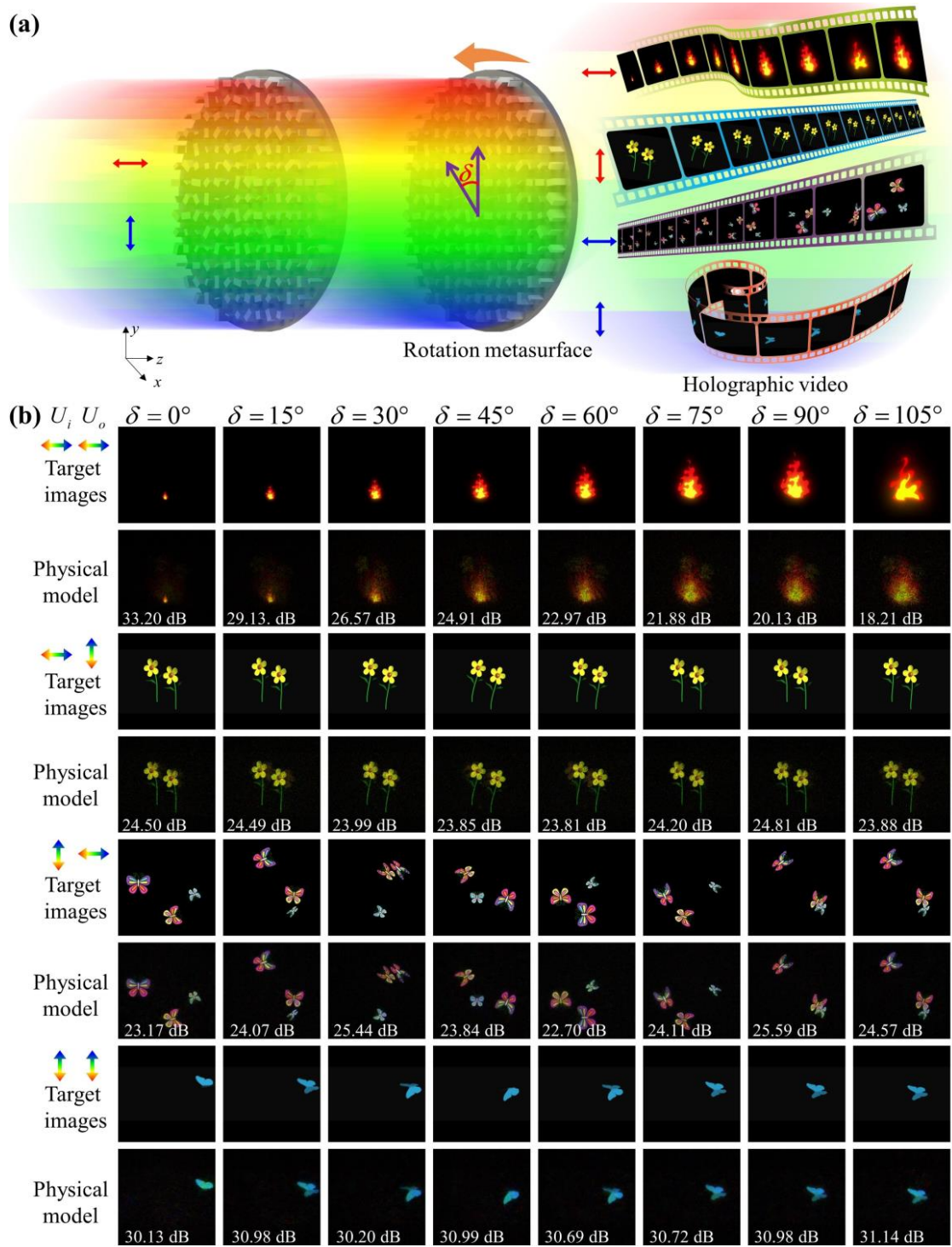

**Figure 5.** Four-polarization-multiplexed and rotationally-tunable metasurface platform for full-color holographic display. (a) Schematic diagram of wavelength, polarization, and rotation multiplexing. (b) Under different input/output polarization, the target and physical model output color holographic images of "Flame", "Yellow flower", "Colorful butterfly", and "Blue butterfly", with the error of each holographic image compared to the target image labeled in the lower edge of the panels. "Flame" by Dao le ge xi gua, CC BY 4.0. https://www.aigei.com/item/mg_dong_hua_huo.html.

Most importantly to note that our end-to-end framework identifies optimal solutions balancing channel quantity and image fidelity under this inherent constraint. Channel proliferation inevitably degrades SNR, though SSIM metrics better align with human visual perception. To reconcile the quality-efficiency trade-off in the meta-diff-DONN model's iterative spatial filtering, we have implemented a strategic compromise: progressively reducing imaging efficiency with channel expansion to preserve visual clarity. Simulation characterization demonstrates a 1.93% optical efficiency while achieving 288-channel multiplexing capability with an average PSNR of 24.77 dB and SSIM of 0.72. Although the achieved transmission efficiency remains comparatively low, which represents a characteristic limitation where increased channel multiplexing substantially degrades performance.[42] Several strategies can counteract this efficiency reduction in multiplexed metasurface-based color holography: (1) Meta-atom pre-screening: Restrict meta-atom selection exclusively to high-transmittance candidates from pre-optimized libraries and refine the DNN fitting to ensure the end-to-end design. (2) Architectural scaling to expand the design space: Enhance wavefront modulation capacity through multilayer integration with increased unit cell density and total number per functional layer. These high-fidelity results confirm that multiplexed channels combining optical parameter encoding and mechanical actuation can be synergistically optimized, demonstrating substantial potential for further system capacity enhancement. The maintained image quality under extreme multiplexing conditions underscores the effectiveness of our co-design paradigm in addressing the fundamental channel-capacity limitations of conventional metasurface holography.

Experimental implementation of rotational cascaded metasurfaces consistently reveals "simulation-to-reality" gaps—discrepancies between computational predictions and experimental outcomes. These deviations primarily stem from manufacturing tolerances, device parameter fluctuations, optical system misalignments, wavefront aberrations, and environmental perturbations, all of which can significantly degrade the final holographic performance (as detailed in Supporting Information S11 on metasurface fabrication defects and alignment errors). To bridge this gap and achieve simulation-experiment consistency and robustness, we propose: (1) Manufacturing defect tolerance: Fabrication imperfections can be modeled as Gaussian noise. Integrating this noise model into the meta-diff-DONN optimization pipeline can help to find metasurface parameters exhibiting inherent immunity to manufacturing variations.[46] However, this strategy becomes challenging to optimize for achieving low-loss convergence when the metasurface involves a large number of multiplexed channels. (2) Misalignment compensation: System misalignments (e.g., metasurface lateral displacement, tilt, or rotation) induce unpredictable wavefront distortions. These distortions are effectively represented by low-order Legendre polynomial combinations, characterized by gradient-continuous phase profiles.[52] Robustness against experimental misalignment is enhanced by superimposing such Legendre-based aberration models during the metasurface inverse design. (3) Enhanced optimization stability: Sensitivity to stochastic perturbations is further reduced by incorporating sharpness-aware minimization into the inverse design framework. This technique constrains the final results to residue in the flat basin in the loss landscape, generally suppressing the loss function (and the device performance) susceptibility against random disturbances.[53,54] The comparison between holograms optimized via end-to-end SAM and those optimized using BP, along with analyses of meta-atom fabrication errors and experimental alignment inaccuracies, are presented in Figures S20–S23. Although SAM training may result in a slight reduction in holographic image quality, it substantially enhances robustness to errors, maintaining high-quality holographic performance even under significant misalignments.

## 4 Conclusion

We propose a general cascaded mechanically reconfigurable metasurface and an end-to-end inverse design framework that multiplexes diverse optical parameters and rotational mechanical operations to achieve multidimensional complex amplitude holographic applications. By developing a differentiable surrogate model for complex amplitude modulation of meta-atoms and targeting multi-channel optical field responses, we directly optimize the metasurface layout through gradient descent algorithms via backpropagation to achieve full optical field control. To highlight its superiority, we demonstrated 18-channel holography by multiplexing wavelength, polarization, and diffraction distance. By multiplexing polarization and rotational mechanical operations, we realized 48-channel dynamic holographic displays, enabling the playback of two monochromatic videos. Furthermore, leveraging wavelength, polarization, and rotational mechanical operations, we achieved high-capacity holographic storage using two cascaded metasurfaces, effectively storing and retrieving 288 relatively high-quality holographic images and dynamically playing 4 color videos.

The framework can accommodate additional multiplexing dimensions, such as other optical field degrees of freedom as mentioned in Figure 1(a). However, it is important to acknowledge inherent limitations in the overall energy efficiency. As channel count and functional complexity increase, fundamental physical constraints—such as structural dispersion and non-independent solution sets of Jones matrices—lead to significant crosstalk and reduced efficiency. To address this, freeform nanostructures could provide a broader design space for exploration, supported by advanced fabrication techniques.[55–57] This framework could further advance intelligent high-throughput holographic display systems, real-time data processing, and automated systems.

## Supporting Information

Supporting Information is available from the Wiley Online Library or from the author.

## Acknowledgments

This work was supported by the National Natural Science Foundation of China (No. 62375064), Shenzhen Science and Technology Program (No. JCYJ20250604145307010 and KJZD20230923114803007), National Key Research and Development Program of China (No. 2022YFB3603204), and Guangdong Basic and Applied Basic Research Foundation (No. 2022A1515011488 and No. 2023A1515110572).

## Conflict of Interest

The authors declare no conflict of interest.

## Data Availability Statement

The data that support the findings of this study are available from the corresponding author upon reasonable request.

**References**

[1] J. Xiong, E.-L. Hsiang, Z. He, T. Zhan, S.-T. Wu, "Augmented Reality and Virtual Reality Displays: Emerging Technologies and Future Perspectives," *Light: Science & Applications*, vol. 10 (2021): 216. https://doi.org/10.1038/s41377-021-00658-8

[2] S. Sun, K.-Y. Yang, C.-M. Wang, et al., "High-Efficiency Broadband Anomalous Reflection by Gradient Meta- Surfaces," *Nano Letters*, vol. 12, no. 12 (2012): 6223. https://doi.org/10.1021/nl3032668

[3] N. Yu, F. Capasso, "Flat Optics with Designer Metasurfaces," *Nature Materials*, vol. 13, no. 2 (2014): 139–150. https://doi.org/10.1038/nmat3839

[4] M. Khorasaninejad, W.T. Chen, R.C. Devlin, J. Oh, A.Y. Zhu, F. Capasso, "Metalenses at Visible Wavelengths: Diffraction-Limited Focusing and Subwavelength Resolution Imaging," *Science*, vol. 352, no. 6290 (2016): 1190–1194. https://doi.org/10.1126/science.aaf6644

[5] R.J. Lin, V.-C. Su, S. W., et al., "Achromatic Metalens Array for Full-Colour Light-Field Imaging," *Nature Nanotechnology*, vol. 14 (2019): 227. https://doi.org/10.1038/s41565-018-0347-0

[6] A.C. Overvig, S. Shrestha, S. C. Malek, et al., "Dielectric Metasurfaces for Complete and Independent Control of the Optical Amplitude and Phase," *Light: Science & Applications*, vol. 8 (2019): 92. https://doi.org/10.1038/s41377-019-0201-7

[7] X. Chen, S. Xin, Q. Liu, et al., "Dielectric Metasurface-Assisted Terahertz Sensing: Mechanism, Fabrication, and Multiscenario Applications," *Nanophotonics*, vol. 14, no. 3 (2025): 271. https://doi.org/10.1515/nanoph-2024-0573

[8] S. Nie, I.F. Akyildiz, "Metasurfaces for Multiplexed Communication," *Nature Electronics*, vol. 4, no. 3 (2021): 177–178. https://doi.org/10.1038/s41928-021-00555-3

[9] Y.M. Ning, Q. Ma, Q. Xiao, Z. Gu, T.J. Cui, "Reprogrammable Nonlinear Transmission Controls Using an Information Metasurface," *Advanced Optical Materials*, vol. 12, no. 3 (2024): 2301525. https://doi.org/10.1002/adom.202301525

[10] J. Sang, Y. Yuan, W. Tang, et al., "Coverage Enhancement by Deploying RIS in 5G Commercial Mobile Networks: Field Trials," *IEEE Wireless Communications*, vol. 31 (2024): 172–180. https://doi.org/10.1109/mwc.011.2200356

[11] J.W. You, Q. Ma, Z. Lan, Q. Xiao, N.C. Panoiu, T.J. Cui, "Reprogrammable Plasmonic Topological Insulators with Ultrafast Control," *Nature Communications*, vol. 12 (2021): 5468. https://doi.org/10.1038/s41467-021-25835-6

[12] T. Zhang, X. Zang, Z. Tan, et al., "Integrated Polarization, Distance, and Rotation for Multi-DoF Diffractive Processor and Information Encryption," *Advanced Materials,* vol. 37, no. 35 (2025): 2506222. https://doi.org/10.1002/adma.202506222

[13] X. Ding, Z. Zhao, P. Xie, et al., "Metasurface-Based Optical Logic Operators Driven by Diffractive Neural Networks," *Advanced Materials,* vol. 36, no. 35 (2024): 2308993. https://doi.org/10.1002/adma.202308993

[14] Z. Fan, C. Qian, Y. Jia, et al., "Holographic Multiplexing Metasurface with Twisted Diffractive Neural Network," *Nature Communications*, vol. 15 (2024): 9416. https://doi.org/10.1038/s41467-024-53749-6

[15] J. Chu, N. Cheriere, G. Brennan, et al., "Can Holographic Optical Storage Displace Hard Disk Drives?" *Communications Engineering*, vol. 3 (2024): 79. https://doi.org/10.1038/s44172-024-00225-0

[16] A.H. Dorrah, F. Capasso, "Tunable Structured Light with Flat Optics," *Science,* vol. 376, no. 6591 (2022): eabi6860. https://doi.org/10.1126/science.abi6860

[17] B. Wang, F. Dong, Q.-T. Li, et al., "Visible-Frequency Dielectric Metasurfaces for Multiwavelength Achromatic and Highly Dispersive Holograms," *Nano Letters*, vol. 16, no. 8 (2016): 5235–5240. https://doi.org/10.1021/acs.nanolett.6b02326

[18] Y.-W. Huang, W.T. Chen, W.-Y. Tsai, et al., "Aluminum Plasmonic Multicolor Meta-Hologram," *Nano Letters*, vol. 15, no. 5 (2015): 3122–3127. https://doi.org/10.1021/acs.nanolett.5b00184

[19] Y.Q. Peng, H.P. Lu, D.S. Zhang, L.C. Wang, Z.Y. Ma, J.J. Xiao, "Inverse Design of a Light Nanorouter for a Spatially Multiplexed Optical Filter," *Optics Letters*, vol. 48, no. 23 (2023): 6232. https://doi.org/10.1364/OL.505205

[20] J.P. Balthasar Mueller, N.A. Rubin, R.C. Devlin, B. Groever, F. Capasso, "Metasurface Polarization Optics: Independent Phase Control of Arbitrary Orthogonal States of Polarization," *Physical Review Letters*, vol. 118, no. 11 (2017): 113901.

https://doi.org/10.1103/PhysRevLett.118.113901

[21] Z. Yin, Y. Yao, J. Zhang, J. Gu, "Multi-Dimensional Reconfigurable, Physically Composable Hybrid Diffractive Optical Neural Network," *(Preprint) arXiv* (submitted November 2024). http://arxiv.org/abs/2411.05748

[22] X. Guo, J. Zhong, B. Li, et al., "Full-Color Holographic Display and Encryption with Full-Polarization Degree of Freedom," *Advanced Materials*, vol. 34, no. 3 (2022): 2103192. https://doi.org/10.1002/adma.202103192

[23] J. Xi, J. Shen, M.T. Chow, T. Li, J. Ng, J. Li, "Deep-Learning Assisted Polarization Holograms," *Advanced Optical Materials*, vol. 12, no. 6 (2024): 2202663. https://doi.org/10.1002/adom.202202663

[24] Z. Liu, X. Mu, H. Song, et al., "Dual Jones Matrices Empowered Six Phase Channels Modulation with Single-Layer Monoatomic Metasurfaces," *Laser & Photonics Reviews*, vol. 19, no. 7 (2025): 2401526. https://doi.org/10.1002/lpor.202401526

[25] B. Xiong, Y. Liu, Y. Xu, et al., "Breaking the Limitation of Polarization Multiplexing in Optical Metasurfaces with Engineered Noise," *Science*, vol. 379, no. 6629 (2023): 294–299. https://doi.org/10.1126/science.ade5140

[26] J. Wang, J. Chen, F. Yu, et al., "Unlocking Ultra-High Holographic Information Capacity through Nonorthogonal Polarization Multiplexing," *Nature Communications* 15 (2024): 6284. https://doi.org/10.1038/s41467-024-50586-5

[27] X. Liu, Z. Ma, D. Zhang, Q. Bao, Z. Liu, J. Xiao, "Hypermultiplexed Off-Chip Hologram by On-Chip Integrated Metasurface," *Advanced Optical Materials*, vol. 12, no. 28 (2024): 2401169. https://doi.org/10.1002/adom.202401169

[28] D. Zhang, T. Ma, Y. Peng, X. Liu, J.-J. Xiao, "Spectro-Polarimetric-Depth Imaging by Inverse-Designed Single-Cell Metasurface," *Journal of Lightwave Technology*, vol 42, no.17 (2024): 6003. https://doi.org/10.1109/JLT.2024.3400845

[29] H. Xiong, X. Zhang, P. Li, et al., "3D Multiview Holographic Display with Wide Field of View Based on Metasurface," *Advanced Optical Materials*, vol. 13, no. 5 (2024): 2402504. https://doi.org/10.1002/adom.202402504

[30] H. Ren, X. Fang, J. Jang, J. Bürger, J. Rho, S. A. Maier, "Complex-Amplitude Metasurface-Based Orbital Angular Momentum Holography in Momentum Space," *Nature Nanotechnology*, vol. 15, (2020): 948-955. https://doi.org/10.1038/s41565-020-0768-4

[31] Y.-L. Li, N.-N. Li, D. Wang, et al., "Tunable Liquid Crystal Grating Based Holographic 3D Display System with Wide Viewing Angle and Large Size," *Light: Science & Applications*, vol. 11 (2022): 188. https://doi.org/10.1038/s41377-022-00880-y

[32] C. Chang, X. Ding, D. Wang, et al., "Split Lohmann Computer Holography: Fast Generation of 3D Hologram in Single-Step Diffraction Calculation," *Advanced Photonics Nexus*, vol. 3, no.3 (2024): 036001. https://doi.org/10.1117/1.APN.3.3.036001

[33] C. Liu, Q. Ma, Z.J. Luo, et al., "A Programmable Diffractive Deep Neural Network Based on a Digital-Coding Metasurface Array," *Nature Electronics*, vol. 5, no. 2 (2022): 113–122. https://doi.org/10.1038/s41928-022-00719-9

[34] X.G. Zhang, W.X. Jiang, H.L. Jiang, et al., "An Optically Driven Digital Metasurface for Programming Electromagnetic Functions," *Nature Electronics*, vol. 3, no. 3 (2020): 165–171. https://doi.org/10.1038/s41928-020-0380-5

[35] B. Chen, S. Yang, J. Chen, et al., "Directional Terahertz Holography with Thermally Active Janus Metasurface," *Light: Science & Applications*, vol. 12 (2023): 136. https://doi.org/10.1038/s41377-023-01177-4

[36] W. Jia, D. Lin, B. Sensale-Rodriguez, "Machine Learning Enables Multi-Degree-of-Freedom Reconfigurable Terahertz Holograms with Cascaded Diffractive Optical Elements," *Advanced Optical Materials*, vol. 11, no. 7 (2023): 2202538. https://doi.org/10.1002/adom.202202538

[37] D. Liao, M. Wang, K.F. Chan, C.H. Chan, H. Wang, "A Deep-Learning Enabled Discrete Dielectric Lens Antenna for Terahertz Reconfigurable Holographic Imaging," *IEEE Antennas and Wireless Propagation Letters*, vol. 21, no. 4 (2022): 823–827. https://doi.org/10.1109/LAWP.2022.3149861

[38] Y. Wang, C. Pang, J. Qi, "3D Reconfigurable Vectorial Holography via a Dual-Layer Hybrid Metasurface Device," *Laser & Photonics Reviews*, vol. 18, no. 2 (2024): 2300832. https://doi.org/10.1002/lpor.202300832

[39] Y. Wang, A. Yu, Y. Cheng, J. Qi, "Matrix Diffractive Deep Neural Networks Merging Polarization into Meta-Devices," *Laser & Photonics Reviews*, vol. 18, no. 2 (2023): 2300903. https://doi.org/10.1002/lpor.202300903

[40] Q. Wei, L. Huang, R. Zhao, et al., "Rotational Multiplexing Method Based on Cascaded Metasurface Holography," *Advanced Optical Materials*, vol. 10, no. 8 (2022): 2102166. https://doi.org/10.1002/adom.202102166

[41] H. Wang, Z. Qin, H. Zhou, et al., "Origami–Kirigami Arts: Achieving Circular Dichroism by Flexible Meta-Film for Electromagnetic Information Encryption," *Laser & Photonics Reviews*, vol. 17, no. 2 (2023): 2200545. https://doi.org/10.1002/lpor.202200545

[42] P. Feng, F. Liu, Y. Liu, et al., "Diffractive Magic Cube Network with Super-high Capacity Enabled by Mechanical Reconfiguration," *(Preprint) arXiv* (submitted December 2024). https://doi.org/10.48550/arXiv.2412.20693

[43] Y. Yin, Q. Jiang, H. Wang, L. Huang, "Color Holographic Display Based on Complex-Amplitude Metasurface," *Laser & Photonics Reviews*, vol. 19, no. 1 (2025): 2400884. https://doi.org/10.1002/lpor.202400884

[44] Y. Yin, Q. Jiang, H. Wang, et al., "Multi-Dimensional Multiplexed Metasurface Holography by Inverse Design," *Advanced Materials*, vol. 36, no. 21 (2024): 2312303. https://doi.org/10.1002/adma.202312303

[45] L. Huang, Z. Han, A. Wirth-Singh, et al., "Broadband Thermal Imaging Using Meta-Optics," *Nature Communications*, vol. 15 (2024): 1662. https://doi.org/10.1038/s41467-024-45904-w

[46] H. Chi, Y. Hu, X. Ou, et al., "Neural Network-Assisted End-to-End Design for Dispersive Full-Parameter Control of Meta-Optics," *Advanced Materials*, vol. 37, no. 13 (2025): 2419621. https://doi.org/10.1002/adma.202419621

[47] T. Ma, X. Liu, Q. Bao, B. Zhang, J.-J. Xiao, "Neural Network-Assisted End-to-End Inverse Design for Polarimetric Microspectrometer," *Advanced Materials Technologies*, vol.10, no. 14 (2025): 2401999. https://doi.org/10.1002/admt.202401999

[48] X. Lin, Y. Rivenson, N.T. Yardimci, et al., "All-Optical Machine Learning Using Diffractive Deep Neural Networks," *Science*, vol. 361, no. 6406 (2018): 1004–1008. https://doi.org/10.1126/science.aat8084

[49] H. Chen, J. Feng, M. Jiang, et al., "Diffractive Deep Neural Networks at Visible Wavelengths," *Engineering*, vol. 7, no. 10 (2021): 1483–1491. https://doi.org/10.1016/j.eng.2020.07.032

[50] T. Yan, J. Wu, T. Zhou, et al., "Fourier-space Diffractive Deep Neural Network," *Physical Review Letters*, vol. 123, no. 2 (2019): 023901. https://doi.org/10.1103/PhysRevLett.123.023901

[51] T.-C. Poon, J.-P. Liu, Introduction to Modern Digital Holography: With Matlab, (Cambridge: Cambridge University Press, 2014), ISBN 9781107016705.

[52] E. Kewei, C. Zhang, M. Li, Z. Xiong, D. Li, "Wavefront Reconstruction Algorithm Based on Legendre Polynomials for Radial Shearing Interferometry over a Square Area and Error Analysis," *Optics Express*, vol. 23, no. 16 (2015): 20267. https://doi.org/10.1364/OE.23.020267

[53] D. Mengu, Y. Rivenson, A. Ozcan, "Scale-, Shift-, and Rotation-Invariant Diffractive Optical Networks," *ACS Photonics*, vol. 8, (2021): 324–334. https://doi.org/10.1021/acsphotonics.0c01583

[54] T. Xu, Z. Luo, S. Liu, et al., "Perfecting Imperfect Physical Neural Networks with Transferable Robustness using Sharpness-Aware Training," *(Preprint) arXiv* (submitted November 2024). https://doi.org/10.48550/arXiv.2411.12352

[55] R.K. Padhy, A. Chandrasekhar, "PhoTOS: Topology Optimization of Photonic Components using a Shape Library," *(Preprint) arXiv* (submitted July 2024). http://arxiv.org/abs/2407.00845

[56] C. Kim, B. Lee, "TORCWA: GPU-Accelerated Fourier Modal Method and Gradient-Based Optimization for Metasurface Design," *Computer Physics Communications*, vol. 282 (2023): 108552. https://doi.org/10.1016/j.cpc.2022.108552

[57] C. Kim, J. Hong, J. Jang, et al., "Freeform Metasurface Color Router for Deep Submicron Pixel Image Sensors," *Science Advances, vol.* 10, no. 22 (2024): eadn9000. https://doi.org/10.1126/sciadv.adn9000